Active Tuning of Surface-Phonon Polariton Resonances *via* Carrier Photoinjection

Adam D. Dunkelberger,[1,*] Chase T. Ellis, [1,*] Daniel C. Ratchford, [1] Alexander J. Giles, [1]
Mijin Kim, [2] Chul S. Kim, [1] Bryan T. Spann,[3,4] Igor Vurgaftman, [1] Joseph G. Tischler, [1]
James P. Long,[1,5] Orest J. Glembocki,[1,5] Jeffrey C. Owrutsky, [1] and Joshua D. Caldwell[1]

1. U.S. Naval Research Laboratory, 4555 Overlook Ave. S.W., Washington, D.C. 20375

2. Sotera Defense Solutions, Inc., Columbia, MD 21046

3. Former NRC/NRL Postdoctoral Fellow (residing at NRL, Washington, D.C.)

4. Current Address: National Institute of Standards and Technology, Applied Physics
Division, Boulder, CO 80305

5. Retired from NRL, Washington, DC

*Contributed equally to this work.

Abstract:

Surface-phonon polaritons (SPhPs) are attractive alternatives to far-infrared plasmonics for sub-diffractional confinement of infrared light. Localized SPhP resonances in semiconductor nanoresonators are very narrow, but that linewidth and the limited extent of the Reststrahlen band inherently limit spectral coverage. To address this limitation, we report active tuning of SPhP resonances in InP and 4H-SiC by photoinjecting free carriers into the nanoresonators, taking advantage of the coupling between the carrier plasma and optical phonons to blue-shift SPhP resonances. We demonstrate state-of-the-art tuning figures of merit upon continuous-wave (CW) excitation (in InP) or pulsed excitation (in 4H-SiC). Lifetime effects cause the tuning to saturate in InP, and carrier redistribution leads to rapid ($< 50\ ps$) recovery of the tuning in 4H-SiC. This work opens the path toward actively tuned nanophotonic devices, such as modulators and beacons, in the infrared and identifies important implications of coupling between electronic and phononic excitations.

The realization that sub-diffractional confinement of light can be achieved through the careful design of polaritonic materials and nanostructures led to the development of the fields of metamaterials and nanophotonics.[1-4] Confining light to the nanoscale greatly enhances light-matter interactions, enabling new applications.[5-14] In the infrared (IR), significant advances can be realized through the development of active, flat and/or compact optics,[15-18] sources,[19] and detectors.[20] Such confinement is typically achieved using surface-plasmon polaritons (SPPs) but can also be accomplished through a broad range of other polariton modes.[21,22] One of the most promising alternatives within the mid-infrared to terahertz spectral range is the surface-phonon polariton (SPhP).[6,23-28]

For both SPPs and SPhPs, propagating and localized modes can be supported, and the analytical theory and formalism describing the conditions for a given optical mode are the same, provided the material information is correctly incorporated into the dielectric function.[24] Through the fabrication of nanoresonators of polar dielectric materials, geometry-dependent localized SPhP resonances can be established within the Reststrahlen band of that material.[23,29] The Reststrahlen band is the spectral region located between the longitudinal (LO) and transverse optical (TO) phonons where the real part of the dielectric function is negative. While localized SPP resonances are broad and lossy due to the short scattering time of the carriers in metals and heavily doped semiconductors, the ~100x longer lifetimes of optic phonons provide SPhPs with an intrinsic advantage with respect to loss.[23,24,30] For localized SPhPs, this reduction in loss results in substantially narrower resonances with quality factors ($Q = \omega_0/\Delta\omega$, where $\omega_0$ is the resonance center frequency and $\Delta\omega$ is the linewidth) that are over an order of magnitude larger than for SPP resonances.[6,23,29]

These high-Q resonances with deeply subwavelength light confinement can provide a better spectral match to the absorptive features of molecules or electrooptic devices, but they also present a disadvantage in terms of spectral coverage. Active tuning, or shifting the SPhP resonances with an electrical or optical stimulus, could potentially allow one to combine narrow, low-loss resonances with broad spectral coverage. Indeed, a key advantage of many IR SPP systems lies in the ability to tune resonances by carefully controlling the plasma frequency through modulation of the carrier density.[31,32] In this article, we demonstrate active tuning of SPhP resonances by photoinjecting free carriers into polar dielectric nanoresonators, thereby perturbing the dielectric function and spectrally shifting the SPhP resonances. We report this phenomenon, which is achieved by coupling the carrier plasma to optical lattice vibrations, for two material systems: InP and SiC. For InP, Fourier transform infrared (FTIR) reflectance measurements are used to probe the steady-state behavior of a far-infrared localized SPhP while carriers are photoinjected by an above-bandgap continuous-wave (CW) green laser. For SiC, we use time-resolved IR reflectance spectroscopy to directly monitor SPhP resonance shifts upon pulsed, above-bandgap ultraviolet excitation. We measure substantial resonance shifts, $\delta\omega$, with maximum tuning figures of merit (FOMs) $(\delta\omega/\Delta\omega) > 1$ and find evidence that the dynamics of carrier redistribution and decay are crucial to effective active tuning.

**Coupling between longitudinal optical phonons and carrier plasma**

In polar semiconductors, such as InP and SiC, the LO phonons can couple to the free carrier plasma, effectively modifying the material optical properties. This is widely reported in studies of the LO phonon-plasmon coupling (LOPC) effect,[33,34] which results in a spectral shift and broadening of the LO phonon in Raman spectra[33,35] and a 'softening' of the LO phonon edge of the Reststrahlen band, which has been reported in transient studies of GaN[36] and SiC.[37] In addition, theoretical work has shown that these carrier-induced changes can be highly effective in actively tuning SPhP resonances.[37,38] For example, the frequency-dependent, infrared dielectric permittivity for a polar dielectric material is given by[39]

$$\epsilon\left(\omega, N_j\right) = \epsilon_\infty \left(1 + \frac{\omega_{LO}^2 - \omega_{TO}^2}{\omega_{TO}^2 - \omega^2 - i\omega\gamma}\right) - \sum_{j=e,h} \frac{N_j e^2}{\epsilon_0 m_j^* m_0} \frac{1}{\omega^2 + i\omega\Gamma} \quad , \qquad (1)$$

where the first term is due to the LO and TO phonon contribution, and the second term is the Drude, free-carrier contribution that arises from significant electron $\left(N_{j=e}\right)$ and hole $\left(N_{j=h}\right)$ concentrations (a detailed discussion of Eq. (1) is included in the Supplementary Information). The SPhP resonance frequency of a nanoparticle is determined by the condition $Re\left[\epsilon(\omega_{res}, N_j)\right] = \epsilon_R$, where $\epsilon_R$ is the resonant value of the permittivity that depends on the particle shape and the surrounding medium. As such, this condition combined with Eq. (1) indicates that an increase in the electron/hole concentration $(N_j)$ decreases $\epsilon$ at a given frequency, leading to a lower (more negative) value of $Re\left[\epsilon(\omega, N_j)\right]$ at all frequencies throughout the Reststrahlen band and consequently an increase in the resonance frequency required to match $\epsilon_R$. Therefore, the SPhP resonance frequency $\omega_{res}$ blue shifts if carriers are added to the system, as recently demonstrated by Karakachian, *et al*.[40]

**Steady-State Spectroscopy of SPhP resonances**

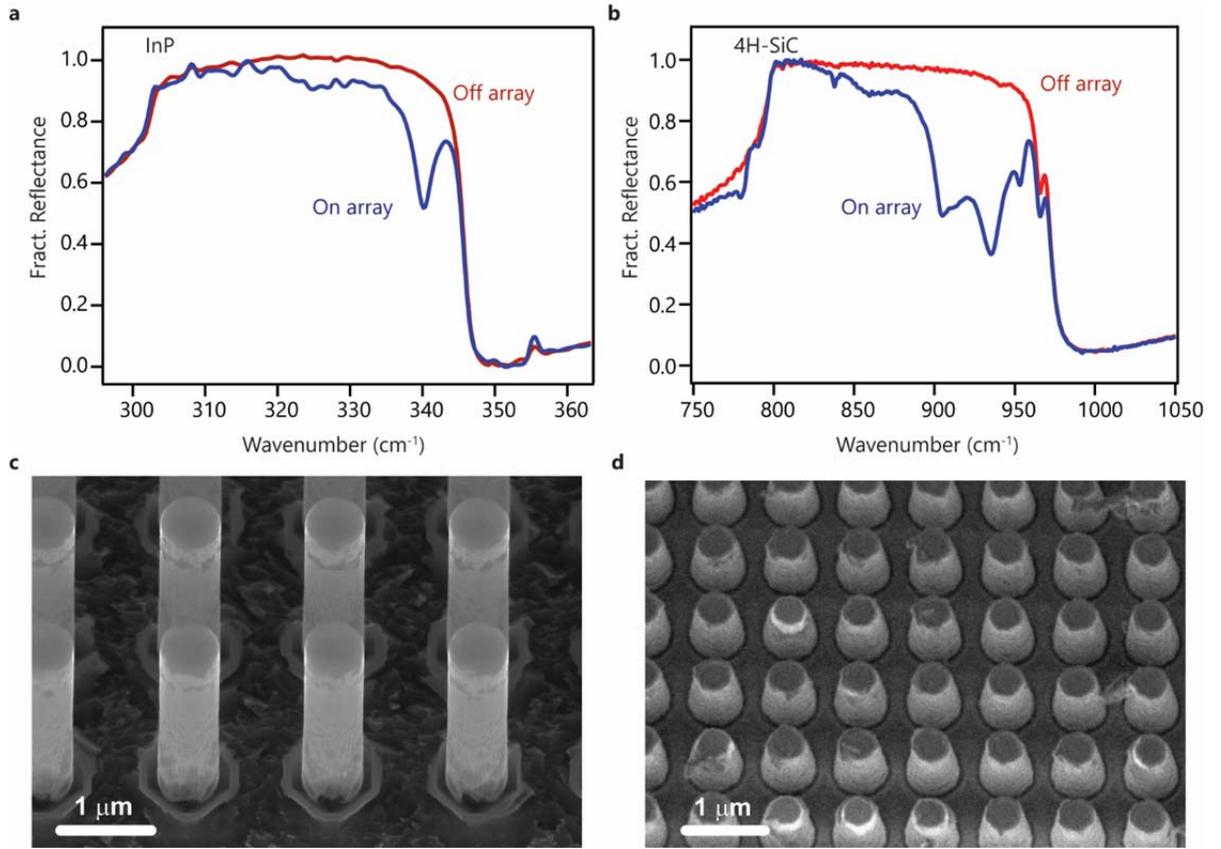

**Figure 1 | Steady-state Spectroscopy of InP and SiC Arrays**. Infrared reflectance spectra of **a** InP and **b** 4H-SiC substrates (red trace), showing the Reststrahlen band between 306-343 cm⁻¹ and 797-972 cm⁻¹, respectively, and nanopillar arrays (blue). Scanning electron micrograph images of **c** InP and **d** SiC nanopillars. The InP sample consists of nanopillars with h = 1500 nm, d = 500 nm, and p = 1500 nm, showing a dipole SPhP resonance at ~340 cm⁻¹, while the 4H-SiC pillars have h = 600 nm, d = 300 nm, and p = 700 nm, showing a monopole SPhP resonance at 900 cm⁻¹ and a transverse dipole SPhP resonance at 936 cm⁻¹, with h, d and p referring to the nanopillar height, diameter and pitch.

Examples of narrow localized SPhP resonances are presented in **Fig. 1** via the infrared reflection spectra of periodic arrays of nanopillars etched into a) an undoped InP substrate and b) an ~160 μm-thick, low-doped (~ $10^{14}$ cm⁻³) SiC epitaxial layer on a highly doped (~ 3 × $10^{18}$ cm⁻³) substrate. The reflectance spectra of unpatterned (red curves) and patterned (blue curves) regions of an undoped InP substrate (**Fig. 1a**), reveal the far-IR Reststrahlen band ($\omega =$

306-343 cm$^{-1}$) and corresponding localized SPhP resonances, respectively. The patterned InP reflectance spectrum, shown for an array consisting of cylindrical nanopillars with height $h = 1500$ nm, diameter $d = 500$ nm, and pitch $p = 1500$ nm (SEM in **Fig. 1c**) exhibits a strongly absorptive, localized SPhP resonance within the Reststrahlen band centered at 340 cm$^{-1}$ with a full width at half maximum (FWHM) of 3.6 cm$^{-1}$. Three-dimensional electromagnetic simulations suggest that this mode corresponds to a transverse (in-plane) dipole resonance (see supplementary information). The corresponding spectra for the unpatterned and patterned regions on the 4H-SiC sample are presented in **Fig. 1b**. The highly reflective Reststrahlen band ($\omega = 797$-972 cm$^{-1}$) along with the localized SPhP resonances are also observed here, with the latter collected from a periodic array of cylindrical nanopillars with $h = 600$ nm, $d = 300$ nm and $p = 700$ nm (SEM in **Fig. 1d**). Prior work establishes that the smaller feature at ~ 900 cm$^{-1}$ is a substrate-modified longitudinal dipole resonance, referred to as the 'monopole,' and the feature at 936 cm$^{-1}$ is a transverse dipole resonance.[23,29,41] It should be noted that a weak feature similar to the monopole is observed near 325 cm$^{-1}$ in the InP spectra (see Supplementary Information).

**Active Tuning in InP Nanopillars**

To investigate the active tuning of SPhP modes in InP nanopillars, we use FTIR measurements to probe resonances as electron-hole pairs are generated by CW above-gap ($E_g = 1.3$ eV; 900 nm) laser illumination ($E_{laser} = 2.33$ eV; 532 nm). The CW laser induces a steady-state concentration of photoinjected carriers, allowing us to explore the influence of the plasma effect upon the narrow localized SPhP resonances. This photoinjection into the same InP nanopillar array shown in **Fig. 1a** results in a non-monotonic shift of the localized SPhP resonance for CW pump intensities ($P$) in the range of 1.3 mW/cm$^2$ – 750 W/cm$^2$ (**Fig. 2a and b**). Within this range, a maximum spectral blue-shift ($\delta\omega$) of 1.4 cm$^{-1}$ is observed for the 3.3 cm$^{-1}$ FWHM ($\Delta\omega$) resonance, corresponding to a FOM ($\delta\omega/\Delta\omega$) of ~0.4. The 1.4 cm$^{-1}$ spectral blueshift is expected to arise from an injected carrier density of $3.2 \times 10^{16}$ cm$^{-1}$ (see supplementary information for details). This is consistent with previous studies of bulk InP that report a comparable shift (~1 cm$^{-1}$) for the LO phonon mode, measured via Raman spectroscopy, with a carrier concentration of $3 \times 10^{16}$ cm$^{-1}$.[42]

As noted above, without laser illumination, we observe a primary transverse dipole resonance located near 340 cm⁻¹, which is analogous to the more extensively studied 4H-SiC system.[23,29,41] The free carriers perturb the permittivity $\left[\epsilon\left(\omega, N_j\right)\right]$, which blue shifts $\omega_{res}$. As shown in **Figs. 2a** and **2b**, this behavior is observed for pump intensities lower than 66 W/cm². However, at higher injection levels ($P > 66$ W/cm²) $\omega_{res}$ red-shifts, consistent with laser-induced heating effects that red-shift the optical phonon frequencies through lattice expansion[43] (see supplementary information). A thermocouple located underneath the sample confirms that sample heating is significant at these high pump powers, where temperatures up to 100 °C were measured. In this high-injection regime, the temperature-dependent changes to the phononic component of the dielectric function overwhelm the induced free-carrier contribution, so the net result is a spectral red-shift of the localized SPhP resonance.

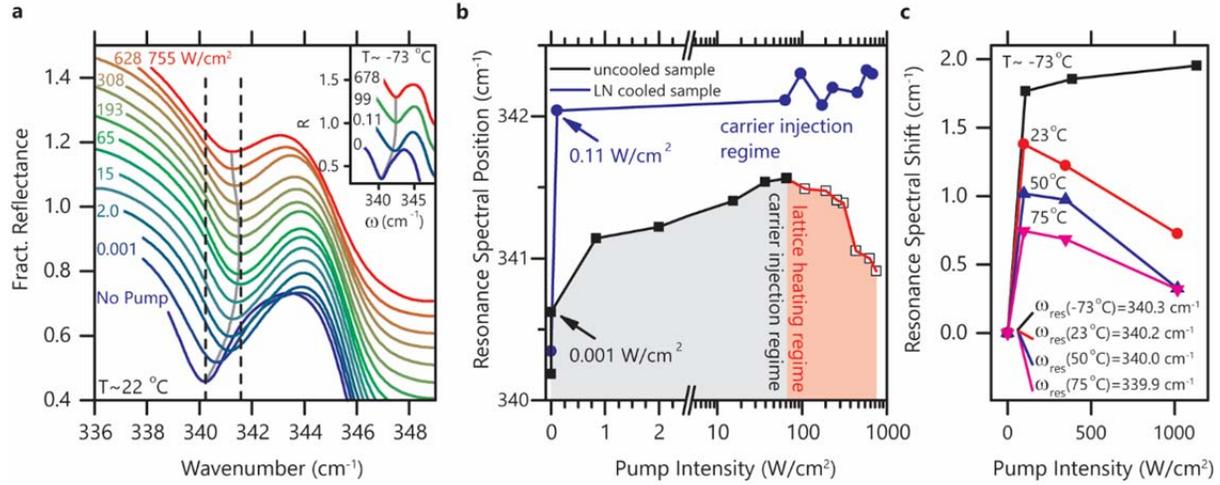

**Figure 2 | Continuous-wave carrier tuning of InP pillar array. a,** Infrared reflectance spectra of InP pillars as a function of incident laser power showing the response of a single localized surface phonon polariton (SPhP) resonance to CW pump laser (λ=532 nm) without regulation of the sample temperature. **b,** The spectral evolution of the localized SPhP resonance spectral position as a function of laser power with (blue circles) and without (black squares) sample cooling. As indicated in both **a** and **b** panels, at low laser powers (P<121 mW), increasing laser power results in a spectral blueshift of the resonance due to carrier injection induced perturbations of the InP dielectric function (black closed squares in panel **b**). However, without thermal regulation of the sample temperature, higher laser powers (P>121 mW) result in significant heating of the sample, yielding a large redshift of the resonance (open black squares in panel **b**). Such heating effects can be suppressed by cooling the sample, as demonstrated by the inset in **a** and blue circles in **b**. In addition to suppressing heating effects, sample cooling also results in larger overall spectral shifts of resonances, as indicated in panel **c**.

To overcome these thermal effects, the sample was cooled with liquid nitrogen (LN) vapor ($T \sim -73\,°C$). Under LN cooling (blue circles, **Fig. 2b**), there is a larger spectral blue-shift of the resonance that appears to saturate at higher pump intensities ($P > 0.11$ W/cm$^2$) with no apparent red-shift, even at the highest injection levels. Furthermore, the onset of the blue-shift is faster with increasing injection at lower sample temperatures, as demonstrated in **Fig. 2c**, in which the pump-intensity dependence of the localized SPhP resonance is presented for four different temperatures. The elevated temperatures are accessed with a heater stage during the photoinjection experiments (red, blue and magenta curves in **Fig. 2c**). Reducing the sample temperature from 75 °C to -73 °C increased the maximum blue-shift by a factor of 2.5 [$\delta\omega_{res}(T = 75\,°C) \approx 0.75\ cm^{-1}$ and $\delta\omega_{res}(T \sim -73\,°C) \approx 1.9\ cm^{-1}$]. This blue-shift, which is on the order of 5% of the Reststrahlen bandwidth, corresponds to a tuning FOM of $\approx 0.57$. Larger shifts are observed for nanopillars with smaller diameters comparable to the $\lambda = 532$ nm pump laser absorption depth (~200 nm). For example, spectral shifts as large as 3.6 cm$^{-1}$ were observed for 150 nm diameter pillars with a height of 1.5 μm (~10% of the Reststrahlen bandwidth) without any cooling, giving an observed FOM of 0.30 (see supplementary information).

The saturation of the resonance shift and temperature dependence of the maximum spectral shifts are unexpected and not fully understood. In this CW-excitation experiment, the steady-state carrier concentration depends on the product of the effective carrier lifetime ($\tau_{eff}$) and the incident pump intensity ($\Delta\omega \propto N \propto P\tau$). For low temperatures, shown in **Fig. 2b**, $\delta\omega_{res}/\delta P$ is large when $P < 0.11$ W/cm$^2$. As such, the low photon flux density in this range requires $\tau_{eff} \geq 2\ \mu s$ to achieve the required carrier density (~$3 \times 10^{16}\ cm^{-3}$) that yields $\delta = 1.9\ cm^{-1}$. This long $\tau_{eff}$ is consistent with carrier relaxation mechanisms dominated by radiative recombination, in which photon reabsorption can lead to the μs lifetimes in bulk, undoped InP.[44] However, it is not clear why the resonance shift in this regime is not limited primarily by the ambipolar carrier diffusion out of the pillars, which should occur on a time scale of $\sim 10$ ns, or by surface recombination.

In the high-intensity regime, we find that $\delta\omega/\delta P$ is small or negative ($P > 0.11$ W/cm$^2$, for the low-temperature data shown in **Fig. 2b**), indicating a dramatic decrease in $\tau$ compared to the low-pump-intensity regime. It is expected that $\tau$ should decrease with increasing $N$ due to reduced effective radiative recombination lifetime,[44,45] suppressed photon reabsorption,[46] reduced surface band bending that increases surface recombination, and increased carrier diffusion out of the pillars.[44,47] In addition, localized pillar heating may also contribute to the observed saturation, most likely because of the temperature-dependent shift of the phonon energies. On the other hand, the sublinear dependence of the carrier density on pump intensity due to faster radiative recombination and diffusion is expected to become pronounced only for carrier densities in the $10^{17}$ cm$^{-3}$ range, in agreement with previous studies on the steady-state photoinjection of carriers into undoped and low-doped bulk InP.[48,49] This suggests that the nanostructured InP surface plays a pivotal role in the carrier dynamics and the maximum photoexcited spectral shift of localized SPhP resonances. Further studies are needed to fully clarify the origin of the observed spectral saturation.

**Active Tuning in 4H-SiC Nanopillars**

In contrast to InP, the Reststrahlen band of 4H-SiC coincides with a region of the infrared spectrum where ultrafast pulses are readily available. This has enabled us to not only observe the carrier-induced tuning of localized SPhP resonances, but also to directly investigate the time-evolution of the resonance shifts using transient IR reflection after free carriers are generated by above-bandgap $\left(E_g = 3.23 \text{ eV}; \sim384 \text{ nm}\right)$ UV ultrafast laser pulses ($E_{laser} = 4.66$ eV; 266 nm; $\sim 200$ fs). These measurements are carried out on the nanopillar array with the IR reflectance spectrum depicted in **Fig. 1b**. Since the UV pump wavelength is significantly shorter than the indirect 4H-SiC bandgap, the penetration depth of the excitation pulses is comparable ($\sim$500 nm)[50,51] to the nanopillar diameter (300 nm) and height (600 nm).

Transient reflection spectra of the 4H-SiC nanopillar array as a function of carrier injection level were collected 5 ps after excitation (**Fig. 3a**), showing photocarrier-induced blue-shifts of the SPhP resonances. The excitation pulse energy was varied (0-5 µJ), which, when converted to absorbed photon density, $N_{phot}$, by dividing the number of photons in each pulse by the product of the beam area and the absorption depth, leads to maximum injection levels in the $1.7 - 8.0 \times 10^{18}$ cm$^{-3}$ range (colored traces). As the photon density increases (light red through dark blue traces), the resonances blue-shift and broaden until eventually, at injection levels greater than $8 \times 10^{18}$ cm$^{-3}$, they become indistinguishable from the background reflection of the sample. At $N_{phot} \sim 5 \times 10^{18}$ cm$^{-3}$ the transverse dipole resonance was observed to shift about $\delta\omega \sim 10$ cm$^{-1}$, with a starting and ending linewidth, $\Delta\omega$, of 7 and $\sim 11$ cm$^{-1}$, respectively. The tuning FOM is shift of the transverse dipole resonance plotted against the carrier or photon density used to calculate or measure the spectrum.

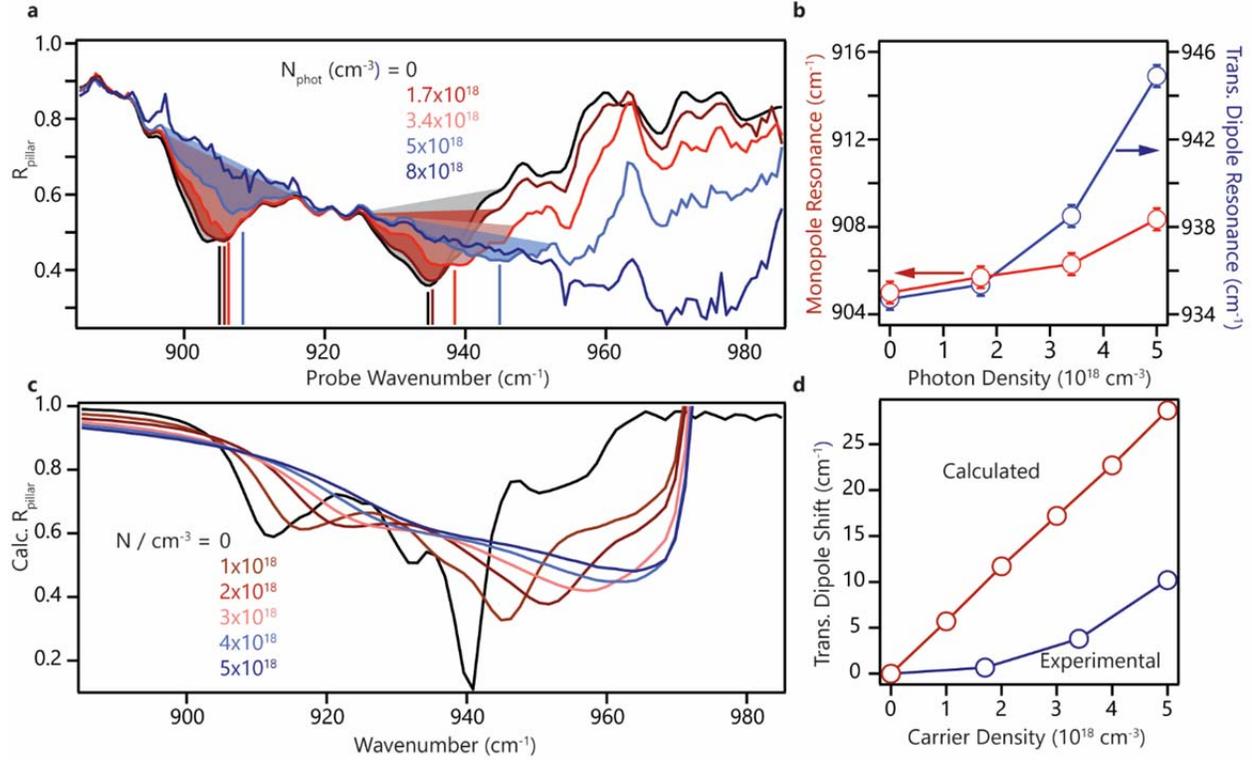

**Figure 3 | Transient Reflection Spectroscopy of SiC Nanopillar Arrays. a**, Transient infrared reflection spectra of 300 nm diameter, 700 nm pitch, 600 nm high 4H-SiC nanopillars on SiC. Traces are reported in terms of R$_{pillar}$ = (R$_{array}$ / R$_{substrate}$)*(R$_{pumped}$ / R$_{array}$) and shown for varying photon densities (N$_{phot}$). The black trace is the unexcited transient spectrum (($\Delta$R / R$_0$) = 0), while the colored traces represent the same but measured 5 ps after a 266 nm (above band gap) excitation. The photon density increases from 1.7x10$^{18}$ cm$^{-3}$ (dark red) to 8x10$^{18}$ cm$^{-3}$ (dark blue). As the incident photon density increases, the SPhP resonances shift to higher frequency and broaden considerably. **b**, Center frequencies of the monopole (red, left axis) and transverse dipole (blue, right axis) resonances for the photon densities whose spectra are reported in **a**. (Note difference y-axis scales.) The 0.5 cm$^{-1}$ error bars are the estimated uncertainties of the transient spectrum. **c**, Transient infrared reflection spectra calculated with finite-difference time domain methods. The black trace is the calculated spectrum for an undoped (N$_{carrier}$ = 0) nanopillar array. Carrier density increases from 1x10$^{18}$ cm$^{-3}$ (dark red) to 5x10$^{18}$ cm$^{-3}$ (dark blue). **d**, Calculated (red) and experimentally measured (blue) magnitude of the resonance-frequency

between 0.9 to 1.4, depending on the linewidth used, offering substantial improvements over other recently reported results for SPhP resonances tuned by laser-induced phase changes in the surrounding medium[52] and other reports of active tuning of localized SPP resonances.[52-56] The one exception is the FOM of 1.8 observed in graphene plasmon resonant structures, albeit with a substantially larger linewidth of ~1100 cm$^{-1}$.[54] The potential combination of SPP and SPhP tuning in concert could allow active tuning in significantly different operational spaces.

In contrast to the transverse dipole, the spectral shift of the monopole resonance was considerably less, with a maximum shift of ~3 cm$^{-1}$, as shown in **Fig. 3b**. This reduction is anticipated because $Re[\epsilon]$ becomes more negative at longer wavelengths as the TO phonon frequency is approached, and so changes in $Re[\epsilon]$ of the same magnitude will have a smaller relative impact on $Re[\epsilon]$ near the monopole than near the transverse dipole resonance. Nevertheless, a tuning FOM of ~0.5 is observed for the monopole, which again compares favorably to the state-of-the-art.

The magnitude of the resonance shifts and the associated broadening of the SPhP resonances with increasing photoinjected carrier density are qualitatively reproduced by finite-difference time-domain (FDTD) simulations (**Fig 3c**). For the calculated spectra, the 4H-SiC dielectric function was taken from Spann et al.[37] with the free-carrier damping time fixed at 12.5 fs. The experimental and calculated spectral shifts of the transverse dipole resonance with carrier concentration are provided in **Fig. 3d**, under the assumption that each absorbed photon excites an electron-hole pair so that the carrier concentration is $N_{phot}$. The calculated shifts increase linearly with carrier concentration, and the predicted shifts are larger than those measured experimentally. We surmise that the discrepancy arises because the FDTD computations assumed a uniform photocarrier distribution, whereas there is a spatial heterogeneity in the carrier distribution within the nanopillars for the experimental work. Again, this results from the shallow penetration depth of the 266 nm excitation pulses (~500 nm) with respect to the nanopillar height (600 nm) and diameter (300 nm), resulting in a spatially exponentially decaying, non-uniform initial carrier density distribution within the nanopillar that does not diffuse in the 5 ps between the excitation and probe pulses. This may reduce the observed spectral shift and a broadening of the resonance. In addition, the actual carrier injection efficiency is likely somewhat below 100%, contributing to the discrepancy between experiment and theory.

The modulation rate of the carrier-induced tuning of SPhP resonances can be an important consideration for some applications. By examining the time dependence of the photoinduced shifts, the limits of the modulation rate can be determined. The time dependence can also provide insight into the processes that limit the tuning range and tuning FOM. The decay of the differential reflection probed at 936 cm$^{-1}$, the transverse dipole resonance frequency, is presented in **Fig. 4** at four different photoinjection levels (blue and red curves). For comparison, we also plot the recovery of the response measured off the array, *i.e.* from the bulk SiC (black curve). Off the array, the response is negative, indicating a lower reflectivity consistent with Reststrahlen-band softening.[37] For both nanostructure resonances and bulk SiC, a portion of the response was observed to recover exponentially in ~ 140 ps, after which the remainder of the signal persists for > 500 ps. Prior work has shown that the bulk carrier recombination lifetime is dominated by Shockley-Read-Hall processes with a lifetime of 900 ns.[50] Because the 140 ps component we observe is independent of excitation pulse intensity and observed from both the patterned and unpatterned surfaces (both etched in the reactive-ion process), it likely arises from surface recombination or from high densities of sub-surface defects that are preferentially probed due to the short excitation laser wavelength.

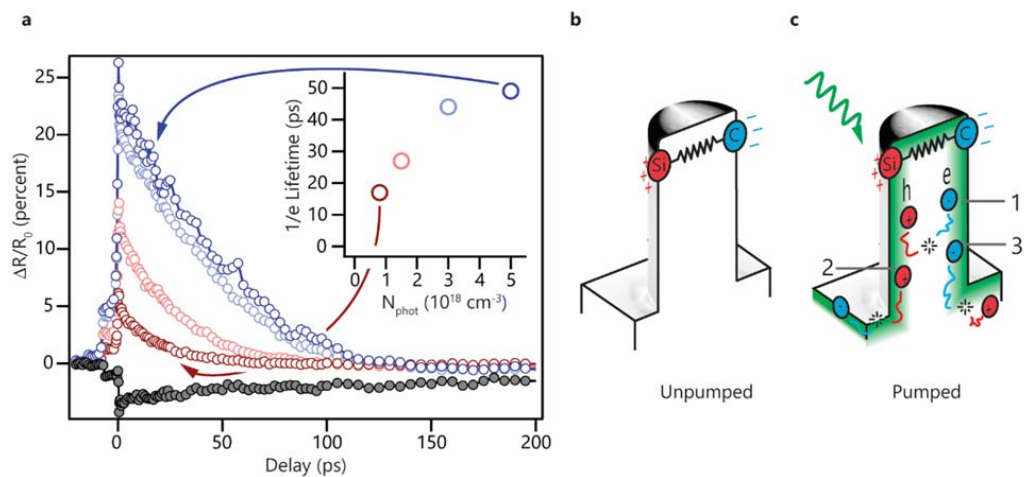

**Figure 4 | Time-Dependence of Resonance Recovery of 4H-SiC Nanopillar Array. a.** Kinetic traces showing differential reflection reported in units of $\Delta R/R_0$ in percent. All traces represent the differential reflection probed at 936 cm$^{-1}$, coincident with the transverse dipole SPhP

resonance, with varying delay time with respect to the 266 nm excitation pulse. The black trace is the response of the unpatterned region of the 4H-SiC sample. Colored traces are the response from the nanopillar array. Photon density increases from 0.8x10$^{18}$ cm$^{-3}$ (dark red) to 5x10$^{18}$ cm$^{-3}$ (dark blue). Inset shows the 1/e lifetime of the transient response for each on-array trace plotted against the incident photon density. **b.** and **c.** schematic of charge dynamics with (panel **b.**) and without (panel **c.**) carrier injection, where the latter shows three relaxation mechanisms for the photoexcited charge: 1) relaxation within the pillar, 2) surface recombination, and 3) diffusion and relaxation within the bulk.

In contrast to the unpatterned regions of the SiC sample, the nanopillar arrays (colored traces in **Fig. 4**) exhibit a positive differential reflectance when probed at the center of the transverse dipole resonance (936 cm$^{-1}$), indicating an increase in reflectance under free-carrier injection. This increase in reflection occurs due to the blue-shift of the resonance, detuning it from the probed frequency. In these arrays, the decay of the transient reflectance is non-exponential, yet we determine the effective 1/e lifetime to compare to the bulk. In addition to the 140 ps decay contribution, a much faster decay is observed with a $N_{phot}$-dependent decay time. At the lowest photon injection levels, this fast response decays in 17 ps. With increasing $N_{phot}$, the 1/e time increases to an apparent maximum of ~ 50 ps occurring at $N_{phot} = 5 \times 10^{18}$ cm$^{-3}$. We emphasize that this decay is likely too fast to arise from carrier recombination, especially within an indirect bandgap semiconductor such as SiC. Instead, one potential explanation is fast redistribution of excited carriers out of the local regions of high optical fields associated with the localized SPhP resonances. The initial carrier distribution upon the excitation pulse is highly non-uniform, and the expected ambipolar diffusion distance is in the 200-300 nm range, depending on the initial position of the carriers in the pillar. The diffusion distance is comparable to the pillar diameter, but smaller than its height, suggesting that most of the redistribution is from the top of the pillar down as well as out of the pillar in the region near its base.

Our results suggest that in a few tens of ps after excitation, spatial redistribution of free carriers occurs, with some of the charges diffusing out of the spatial regions associated with highly SPhP fields. **Figures 4b and 4c** illustrate this redistribution schematically. Insofar as the resonance shift is sensitive to the local permittivity within the nanopillar itself, this redistribution induces a local reduction in the carrier density, resulting in a recovery of the original $\omega_{res}$. If confirmed to limit the carrier dynamics, this diffusion mechanism can be suppressed by increasing the diameter of the nanopillars or by defining isolated pillars on a hetero-substrate with a large conduction-band barrier. After excitation with 300 nm and 325 nm pulses, the SPhP resonances did not shift appreciably. Further experiments on nanopillars with varying geometries are needed to explain this surprising result.

**Discussion**

Active tuning of SPhP resonances in both InP and 4H-SiC clearly depends strongly on the carrier dynamics within the nanopillar arrays. In contrast to InP, where saturation is observed for high excitation powers, the resonances in the 4H-SiC appear to continuously shift until the softening of the Reststrahlen band occludes them. This result suggests that the temperature-dependent effects observed in InP are less important for the 4H-SiC experiment, where the energy deposited by an ultrafast pulse dissipates before the next pulse arrives, perhaps reducing effects from laser heating. However, the smaller effective mass and the lower-frequency LO phonon in InP enables such free-carrier induced spectral shifts to occur with much lower carrier injection levels, which can be realized with off-the-shelf CW lasers.

In summary, our results demonstrate active modulation of SPhP resonances in nanostructured InP and 4H-SiC using continuous-wave and pulsed excitation. Here, an analog the longitudinal optical phonon-plasmon coupling effect that involves the coupling of photoinjected carriers and optical phonons is utilized to alter the permittivity and tune the SPhP bands. We find that this methodology allows large tuning FOMs between 0.5 and 1.5, which are comparable to the state-of-the-art in modulated nanophotonics. In addition, ultrafast experiments on the 4H-SiC nanopillar arrays show that the active tuning can occur on picosecond timescales, even within indirect bandgap semiconductors with much longer recombination lifetimes, and is rapidly reversible. Active tuning of these narrow SPhP resonances has important implications for device applications in modulated optical sources, beacons, modulators, chemical sensors, and other areas. This work points toward many attractive opportunities involving coupling between phonon polaritons and other excitations, especially plasmon polaritons, which are the basis for novel approaches towards active metamaterials and nanophotonics, e.g. the creation of electromagnetic hybrids.[57]

**Acknowledgements**

We acknowledge Dr. Mario Ancona for useful discussions. A.D.D., C.T.E. and A.J.G. acknowledge support from the National Research Council (NRC) – NRL Postdoctoral Fellowship Program.  This work was funded via the Office of Naval Research through the Nanoscience Institute at the U.S. Naval Research Laboratory.

The authors declare no competing financial interests.

**Author Contributions**

A.D.D., C.T.E. J.C.O, J.G.T., J.P.L., O.J.G. and J.D.C. devised the experiment; A.J.G., M.K., C.S.K. fabricated the samples; A.D.D. and B.T.S. carried out the transient reflectance measurements on SiC nanopillars and D.C.R. modeled the results; C.T.E. and J.G.T. carried out the steady-state reflectance measurements on InP nanopillars and I.V. modeled the results; C.T.E. and O.J.G. performed the finite element method simulations. The project was supervised by J.G.T, J.C.O., and J.D.C.

**Methods**

*Fabrication of 4H-SiC nanopillars*

For the experiments discussed here, multiple 600 μm arrays of 4H-SiC nanopillars with varying diameter and pitch were fabricated. This was performed through standard electron beam lithography using a bilayer PMMA resist and reactive ion etching. The pillars were etched to a height of ~ 600 nm, resulting in cylindrical nanopillars of ~ 300 nm diameter spaced on a 700 nm center-to-center pitch. A full description of the fabrication procedures can be found elsewhere.[29] These structures were fabricated into an ~160 μm thick, low doped (~ $1 \times 10^{14}$ cm$^{-3}$) epilayer grown on an 8° off-cut, highly doped ($3 \times 10^{18}$ cm$^{-3}$) substrate, the same substrate studied in Ref. 36.

*Fabrication of InP nanopillars*

Multiple $300 \times 300$ μm nanopillar arrays with varying diameter and pitch were fabricated on undoped InP (background carrier concentration $< 1 \times 10^{16}$ cm$^{-3}$, *n*-type). Fabrication consists of producing a SiN hard mask that defines the pillar cross-section using electron beam lithography and fluorine-based inductively coupled plasma etching (ICP). The pillars were etched into the InP via a chlorine-based ICP. The resulting InP nanopillars discussed in this study have a diameter of 500 nm, height of 1500 nm, and a center-to-center pitch of 1500 nm. The nanopillars discussed in the supplementary information have a diameter of 150 nm, height of 300 nm, and center-to-center pitch of 300 nm.

*Ultrafast Transient Reflection Spectroscopy*

A regenerative amplifier (Coherent Legend) generates 2 mJ, 100 fs pulses centered at 800 nm. Successive BBO crystals frequency triple a portion of the laser fundamental to produce the 266 nm excitation pulses. An optical parametric amplifier (Light Conversion TOPAS) converts the remainder of the laser fundamental to signal and idler pulses in the near infrared and a GaSe crystal converts the signal and idler to the mid infrared *via* difference frequency generation, yielding the 10-12 μm probe pulses. The probe pulses traverse a computer controlled delay stage, after which ZnSe lenses focus the probe pulses onto the sample and collect the probe light reflected at ca. 12° from normal. We do not focus the slightly elliptical ($D_A = 1.1$ mm, $D_B = 1.5$ mm) excitation pulses after the sample. After interacting with the sample, the probe pulses are directed to a monochromator with ~3 cm$^{-1}$ resolution and single-element mercury-cadmium-telluride (MCT) detector. Box-car averagers and lock-in amplifiers synchronized with an optical chopper in the excitation beam measure the excitation-induced differential reflection from the sample. We report the transient response in terms of R$_{pillar}$, which we calculate in two steps. We first measure the unexcited spectrum, R$_0$, by dividing the spectrum reflected from the nanopillar array by the spectrum reflected by the unpatterned surface. Then, we multiply this R$_0$ by the R/R$_0$ response we obtain from the pump-probe measurement to obtain R$_{pillar}$. In general, we measure the reflection from the sample at a specific time delay with respect to the excitation pulse and compare it to the reflection without the excitation pulse. LabVIEW software drives the delay stage and collects data from the lock-in amplifiers. We estimate 10% uncertainty in the 1/e times extracted based on the statistics obtained from averaging multiple scans.

*Steady-state Reflection Spectroscopy*

Steady-state reflectance measurements are carried out using a Bruker 80v spectrometer that is coupled to an all-reflective, infrared microscope (Bruker Hyperion 1000). The microscope is fit with a reverse-Cassegrain microscope objective that focuses unpolarized infrared light onto the sample with an angle of incidence that varies from $12° - 23.6°$. Light reflected by the sample is collected through the same microscope objective and subsequently focused onto and detected by a HgCdTe detector. Steady-state reflectance measurements are normalized to the reflectance of a gold mirror. Samples are pumped with a $\lambda = 532 \ nm$ diode pumped solid state laser (Lighthouse Photonics, Sprout), which is focused down to a spot-size of $450 \ \mu m$ and the measurement probe area is $300 \ \mu m \ \times 300 \ \mu m$. Temperature-dependent measurements, for $T > 22 \ °C$, are carried out with a windowless, heater stage. Low-temperature measurements are carried out with a custom-built, window-less, flow cryostat that is cooled by cold nitrogen gas that passes through a liquid nitrogen heat exchanger.

538